\documentclass[
prb,
]{revtex4-2}

\usepackage{stfloats}
\usepackage{amssymb}
\usepackage{amsmath}

\usepackage{array}
\usepackage{adjustbox}
\usepackage{float}
\usepackage{lipsum, babel}

\usepackage{mathtools}  
\mathtoolsset{showonlyrefs}  
\usepackage{graphicx}% Include figure files
\usepackage{dcolumn}% Align table columns on decimal point
\usepackage{bm} % bold math
\usepackage{subfig}

\begin{document}

\preprint{APS/123-QED}

\title{A3SA: Advanced Data Augmentation via Adjoint Sensitivity Analysis}% Force line breaks with \\
% Photonic data Augmentation via Adjoint Sensitivity Analysis

\author{Chanik Kang\dag}
\affiliation{Department of Mechanical Engineering, Massachusetts Institute of Technology (MIT), Cambridge, MA, 02139, USA}
\affiliation{Now, Department of Artificial Intelligence, Hanyang University, Seoul, 04763, South Korea}

\author{Dongjin Seo\dag}
\affiliation{Department of Electronic Engineering, Hanyang University, Seoul, 04763, South Korea}

\author{Svetlana V. Boriskina}
\affiliation{Department of Mechanical Engineering, Massachusetts Institute of Technology (MIT), Cambridge, MA, 02139, USA}

\author{Haejun Chung*}
\affiliation{\mbox{Department of Electronic Engineering, Hanyang University, Seoul, 04763, South Korea} \\
\mbox{Department of Artificial Intelligence, Hanyang University, Seoul, 04763, South Korea}
\\Email: haejun@hanyang.ac.kr}

\begin{abstract}
% original
% Innovative machine-learning approaches have enabled the inverse design of photonic structures in many practical applications. However, in these approaches, both the number of data and initial data distribution are most important for discovering highly efficient photonic devices, which often require simulated data of thousands to a few hundred thousand. This problem has been a primary bottleneck in machine learning-based photonic design problems. Here, we propose a new data augmentation algorithm based on the adjoint method where it can generate $>$1000$\times$ data with higher device efficiency. The adjoint method predicts the changes of the Figure of Merit (FoM) with respect to a structural perturbation, where this prediction only requires two runs of full-wave Maxwell simulation. Using the adjoint gradient values, we can augment and label a few thousand new data without further computation. Also, the augmented data from the proposed algorithm show a more significant figure of merits because the adjoint gradients provide a highly accurate prediction of FoM changes for the structural variation. We apply the proposed algorithm to a multi-layered metalens design problem. It shows an increase of data generation efficiency by 343 times, and the optimized metalens demonstrates a maximum focusing efficiency of 93\% after applying it to a generative adversarial network (GAN) with the initial dataset of average 18\% efficiency.
 
Innovative machine learning techniques have facilitated the inverse design of photonic structures for numerous practical applications. Nevertheless, within these approaches, the quantity of data and the initial data distribution are paramount for the discovery of highly efficient photonic devices. These devices often require simulated data ranging from thousands to several hundred thousand data points. This issue has consistently posed a major hurdle in machine learning-based photonic design problems. Therefore, we propose a novel data augmentation algorithm grounded in the adjoint method, capable of generating more than 300 times the amount of original data while enhancing device efficiency. The adjoint method forecasts changes in the figure of merit (FoM) resulting from structural perturbations, requiring only two full-wave Maxwell simulations for this prediction. By leveraging the adjoint gradient values, we can augment and label several thousand new data points without any additional computations. Furthermore, the augmented data generated by the proposed algorithm displays significantly improved FoMs owing to the precise FoM change predictions enabled by the adjoint gradients. We apply this algorithm to a multi-layered metalens design problem and demonstrate that it consequently exhibits a 343-fold increase in data generation efficiency. After incorporating the proposed algorithm into a generative adversarial network (GAN), the optimized metalens exhibits a maximum focusing efficiency of 92.93\%, comparable to the theoretical upper bound (93.80\%).
% (1)
% the optimized metalens demonstrates a maximum focusing efficiency of 93\% after applying it to a generative adversarial network (GAN) with the initial dataset of average 18\% efficiency.

%(2)
%After incorporating these data into a generative adversarial network (GAN) alongside the initial dataset, which has an average efficiency of 18\%, the optimized metalens achieved a maximum focusing efficiency of 93\%.

\dag These authors contributed equally to this work.

\end{abstract}

\maketitle

\section{Introduction}

The field of photonics, which involves a study of detection, generation, and manipulation of light, has advanced with the growing interest in its versatile applications including light detection and ranging (LiDAR)~\cite{kim2021nanophotonics, li2022progress, juliano2022metasurface}, optical communication~\cite{kaushal2016optical, khalighi2014survey, hranilovic2006wireless, eldada2004optical}, imaging~\cite{PhysRevLett.85.3966, liu_far, Jacob:06}, optical sensing~\cite{amao2003probes, morales2015nanopaper, photonics8080342, steinegger2020optical, canfarotta2013polymeric, yalcin2006optical}, quantum computing~\cite{slussarenko2019photonic, madsen2022quantum, killoran2019strawberry, gupta2023silicon, takeda2019toward, rudolph2017optimistic, kok2007linear}, and holography~\cite{escuti2004holographic, berger1997photonic, deng2017metasurface, sharp2003holographic, campbell2000fabrication, zito2008two}. In particular, nanophotonics~\cite{saleh2019fundamentals, prasad2004nanophotonics, rigneault2010nanophotonics, so2020deep, ParkKimNamChungParkJang+2022+1809+1845}, which merges the principles of nanotechnology and photonics to control light at the nanoscale, has facilitated the precise implementation of complex photonic structures. Furthermore, the demand for highly efficient and complex structures has led to a need for advanced design techniques.

Conventional photonic design approaches include parameter sweep~\cite{remski2000analysis, mouradian2017rectangular, piggott2018automated}, Bayesian optimization~\cite{gao2022automatic, schneider2019benchmarking, ayassi2021bayesian, sun2023nonlinear, garcia2021bayesian}, and global optimizations such as particle swarm optimization~\cite{kennedy1995particle, chung2020tunable, flannery2018fabry, park2019ultimate, li2019inverse} or genetic algorithms~\cite{holland1992genetic, jafar2018adaptive, huang2019optimization, lee2017concurrent}.  However, these methods encounter significant limitations when confronted with intricately complex design problems. More recently, efficient inverse design approaches have been proposed in photonics~\cite{Miller:EECS-2012-115, molesky2018inverse, su2019nanophotonic, ahn2022photonic, Chung:20, christiansen2021inverse}. Inverse design in photonics is a framework for optimizing photonic structures with respect to the figure of merit (FoM) in a parameter space with many degrees of freedom. One prominent methodology for the inverse photonic design involves the utilization of adjoint sensitivity analysis~\cite{CAO2002171, allaire01242950, Miller:EECS-2012-115}, which predicts the gradient of the FoM with respect to the dielectric permittivity values of engineered materials with only two runs of simulations. The gradient value is utilized to update design parameters to increase the FoM, and this iterative process is referred to as adjoint optimization
% This framework is designed to enhance the optimization of photonic structures concerning a Figure of Merit (FoM) within a vast design parameter space. An established technique for inverse design involves the utilization of adjoint sensitivity analysis ~\cite{CAO2002171, allaire01242950, Miller:EECS-2012-115}, enabling the prediction of the FoM gradient relative to permittivity variations with just two simulation runs
~\cite{Miller:EECS-2012-115, chung2022inverse,wang2018adjoint,jensen2011topology, xiao2016diffractive, piggott2015inverse, mansouree2020multifunctional}. However, the adjoint optimization process often converges to local rather than global optimal designs or requires an intricate binarization process~\cite{gertler2023physical, jiang2019global, jiang2019simulator, jiang2020metanet, fan2020freeform}, which sometimes leads to the degradation of the FoM. 

Meanwhile, deep learning has proven its effectiveness in designing complex photonic structures through the high-level expression of complex and nonlinear functions empowered by data-driven approaches~\cite{Sajedian:19, doi:10.1021/acsphotonics.1c00839, park2023physicsinformed, jiang2019free, 10.1145/3447548.3467414}. In particular, generative models~\cite{jiang2019global, jiang2019free, an2021multifunctional, zhang2023diffusion, 10.1145/3447548.3467414} have piqued considerable interest owing to their power of representation and flexibility in learning the complex structures of given image data. However, deep learning inherently requires a large dataset, often including a few hundred thousand data points, and relies heavily on the initial data distribution~\cite{goodfellow2016deep, van2001art, shorten2019survey, shorten2021text}. Furthermore, conventional generative models~\cite{rezende2016variational, kingma2022autoencoding}  intrinsically do not improve device performance because their optimization function generally relies on the likelihood or its correlated value~\cite{Bond_Taylor_2022}. A generative adversarial network (GAN)~\cite{goodfellow2014generative} represents an innovative approach for training neural networks, which comprises two distinct components: a generator and a discriminator. These two networks engage in a competitive interaction, wherein each strives to outperform the other, leading to mutual improvement. In the context of photonics, GANs exhibit remarkable generation efficiency, enabling rapid generation of numerous photonic devices~\cite{jiang2019free, an2021multifunctional, 10.1145/3447548.3467414}. However,
GANs rarely demonstrate performance enhancements due to their dependence on the distribution of training data. A further complication arises from the loss function used in GAN training, which creates a minimax game between the generator and the discriminator, making it difficult for the learning process to converge~\cite{hsieh2021limits}. Consequently, a new method for utilizing generative models in photonics is required.

In this study, we introduce an innovative data augmentation algorithm for deep-learning-based photonic design. This algorithm, which we name A3SA (Advanced Data Augmentation via Adjoint Sensitivity Analysis), is built on the principle of adjoint sensitivity analysis. It can generate over 1,000 times the initial dataset without requiring many simulations, while simultaneously improving the distribution of the augmented data. Specifically, the adjoint gradients provide a highly accurate prediction of FoM changes caused by structural variations, resulting in augmented data with much higher device efficiencies. Consequently, the A3SA algorithm overcomes the limitations of deep learning in photonics such as the need for a large dataset and reliance on the initial data distribution. In addition, the algorithm can avoid convergence to the local optimum structure, which is often observed in adjoint optimization. We apply the proposed algorithm to a multi-layered metalens design problem involving high structural degrees of freedom. A3SA shows up to 343 $\times$ data augmentations from the original dataset of size 100. In addition, we apply A3SA to a GAN and discover a multi-layered metalens design with a focusing efficiency of 92.93\%. Based on its high versatility and efficiency, our data augmentation algorithm may open a new era in data-driven design in photonics.

\section{Advanced Data Augmentation via Adjoint Sensitivity Analysis}

%We categorize conventional design approaches for optimizing metalens as three strategies: 1) unit-cell approach with conventional optimization methods (e.g., global optimizations), 2) unit-cell approach with data-driven approaches, and 3) adjoint optimization of the entire metalenses. First, the unit-cell method involves segmenting the design space into smaller sections termed `unit cells'~\cite{}. The algorithm effectively reduces the design space from the entire structure to the subarea, which enables a large-scale design with stitching pre-optimized unit cells. 

%However, the unit-cell methods generally are limited to periodic boundary conditions. Also, the unit-cell method assumes field continuity, which implies that the technique can only be utilized in situations with slowly varying phase and amplitude. The process of designing unit cells can be divided into conventional algorithms and data-driven approaches.

%Conventional algorithms such as particle swarm optimization (PSO) or genetic algorithm (GA) 

\begin{figure}[H]
    \centering
    \captionsetup{width=\columnwidth, justification=raggedright}
    \includegraphics[width =0.95\columnwidth]{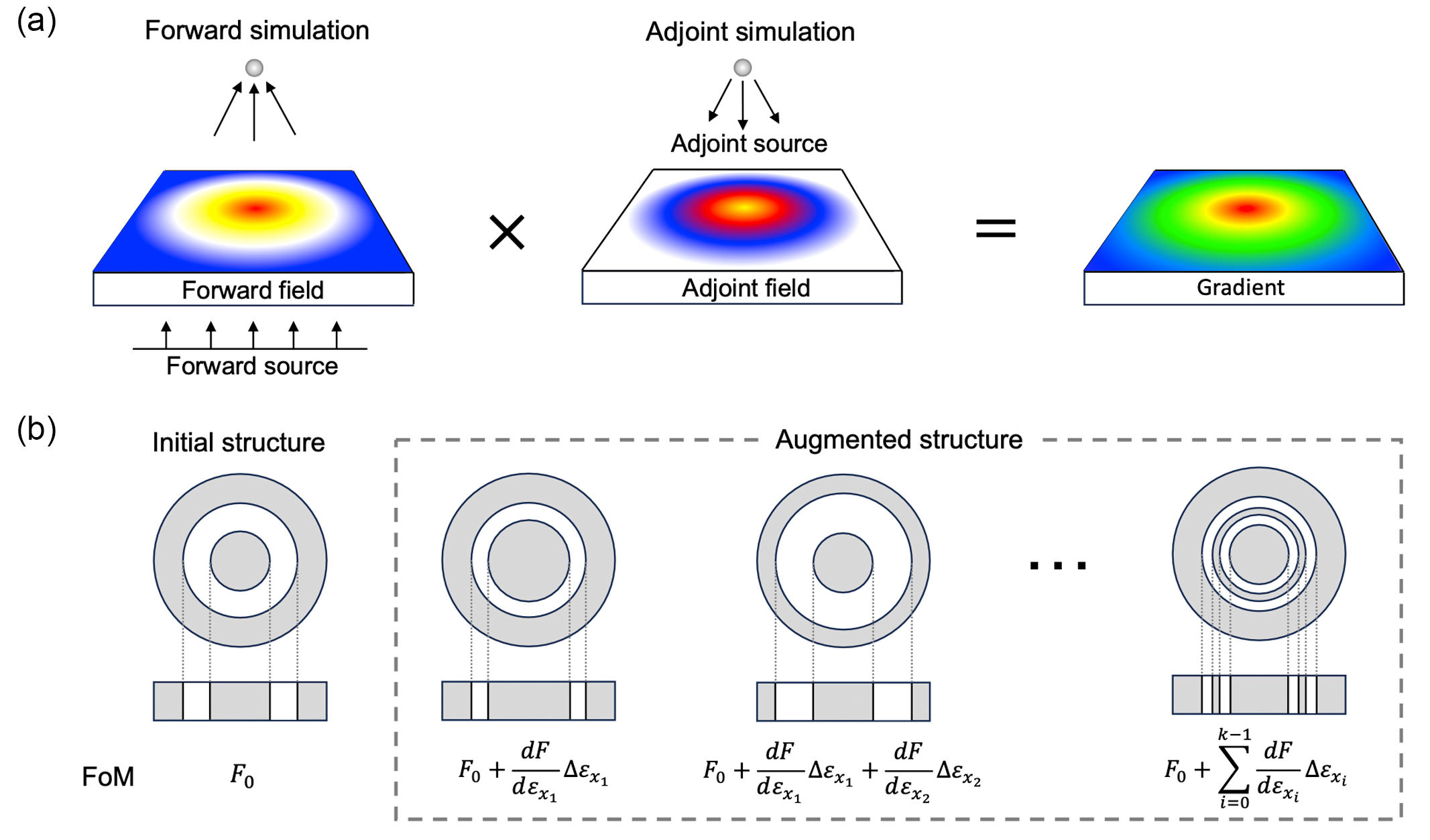}
    
    \caption{Schematics of our A3SA (Advanced Data Augmentation via Adjoint Sensitivity Analysis) algorithm. (a) Illustration of the adjoint sensitivity analysis. Instead of running independent simulations for every geometrical perturbation, the adjoint sensitivity analysis employs the forward and adjoint (backward) simulations to compute the gradients with respect to the numerous geometrical degrees of freedom. Combining the Lorentz reciprocity and Born approximation~\cite{molesky2018inverse, Miller:EECS-2012-115}, the adjoint sensitivity analysis can provide exact gradients ($\frac{dF}{d\varepsilon}$) of the given structure with only two simulations. (b) A3SA algorithm can increase the amount of the initial photonic data with given structural gradients by more than a thousandfold depending on the number of structure parameters. Furtheromore, the additionally-generated photonic data have a FOM distribution shifted to a higher mean value by manually inverting the material density function ($\rho$) using computed gradient information $\frac{dF}{d\varepsilon}$.}
    \label{fig: Figure 1}
\end{figure}

We introduce the A3SA, a novel photonic data augmentation method based on adjoint sensitivity calculations. The calculation provides gradients with respect to the FoM over the design space within only two simulations: forward and adjoint, as illustrated in Fig.~\ref{fig: Figure 1} (a). The critical step in the adjoint sensitivity calculation is the efficient computation of the gradients with respect to the numerous geometrical degrees of freedom by combining the Lorentz reciprocity and Born approximation~\cite{molesky2018inverse, Miller:EECS-2012-115}. Born approximation allows to represent small changes in the dielectric constant by dipole sources with magnitude linearly proportional to an unperturbed field E at the same point. In turn, the reciprocity principle allows obtaining an adjoint field by using coherent dipole sources with amplitudes calculated using the definition of the design FoM, as shown in Fig.~\ref{fig: Figure 1} (a). Then, the variation in the FoM can be calculated as $\mathfrak{Re}\left( \mathbf{E}_{\text{dir}} \cdot \mathbf{E}_{\text{adj}}^* \right)$, where $\mathbf{E}_{\text{dir}}$ and $\mathbf{E}_{\text{adj}}^*$ can be obtained from the forward and adjoint simulations.
The critical insight behind our algorithm is that the computed adjoint gradient value serves as both a ``navigator" and a ``barometer" for newly generated photonic data. Firstly, as a ``navigator", it effectively guides the distribution of the Figure of Merit (FoM) by enhancing it. Secondly, as a ``barometer", it accurately labels the FoM for newly generated data.

%The critical insight behind our algorithm is that the computed adjoint gradient value serves as both a navigator (increased FoM distribution) and a barometer (labeling FoM) for newly generated photonic data. 

The A3SA algorithm starts with a randomly generated initial structure, followed by the computation of adjoint gradients over the design space using the adjoint sensitivity analysis. An example of this process is illustrated in Fig.~\ref{fig: Figure 1} (b) for the case of a photonic structure with cylindrical symmetry. Here, the adjoint gradients are averaged over the smallest design feature with cylindrical symmetry, a width of $50$nm, and a height of $500$nm. A3SA then searches for the highest absolute adjoint gradient ($|g|$). Next, it inverts the material density function ($\rho$) of the smallest design feature having maximum value of $|g|$. The inversion rule is the following: if an adjoint gradient is positive $and$ $\rho$ is negative, an inversion takes place to increase $\rho$ (i.e., inversion from low refractive index to high refractive index). If an adjoint gradient is negative $and$ $\rho$ is positive, an inversion reduces $\rho$ (i.e., inversion from high refractive index to low refractive index).

The new structure with a locally inverted material density function is augmented data with a greater FoM. Material density inversion can be repeated multiple times until the total structural change exceeds the Born approximation in the adjoint sensitivity analysis. Therefore, the amount of augmented data can increase more than a thousandfold in a large photonic design problem, where small local changes do not violate the Born approximation. Augmented photonic structures can also have negligible label error with a FoM of $F+\Delta F$, where $\Delta F \approx \frac{dF}{d\varepsilon}\Delta\varepsilon$, which enables A3SA to be utilized in a score-based deep neural network (DNN) model.
 %This process simplifies the design process since material penalization, one of the hurdles in adjoint optimization, is inherently omitted. 
 The total number of inverted cells per iteration, denoted by $k$, is set as the model hyperparameter. The value of $k$ is proportional to the size of the design area and must not exceed a certain threshold to satisfy the Born approximation validity range. 
The optimal value of $k$ can be determined by performing multiple inversions and analyzing the data distribution. Next, we select the optimal number of inversions by examining the mean and the maximum values of the distribution at each iteration.

In the mathematical context, if we define the design space as $X$ and design feature as $\mathbf{x}$ with each $\mathbf{x}_i$'s corresponding permittivity value represented as $\varepsilon_{\mathbf{x}_i}$, the optimization process of one iteration is expressed in Eq. \eqref{equation_selection}, \eqref{equation_standard}, and \eqref{equation_change}. Equation \eqref{equation_selection} shows the procedure for selecting design features based on their adjoint gradient values, starting from the absolute maximum and proceeding to the next design feature with each adjoint gradient value.
\begin{equation}
\begin{gathered}
    \mathbf{x}_1 = \underset{\mathbf{x}_i \in X}{\mathrm{argmax}} \left| \frac{dF}{d\varepsilon_{\mathbf{x}_i}}  \right|  \\
    \mathbf{x}_2 = \underset{\mathbf{x}_i \in X\setminus \{{\mathbf{x}_1}\} }{\mathrm{argmax}} \left|  \frac{dF}{d\varepsilon_{\mathbf{x}_i}}  \right| \label{equation_selection}\\
    \vdots \\
    \mathbf{x}_k = \underset{\mathbf{x}_i \in X \setminus \{{\mathbf{x}_1, \mathbf{x}_2, ... , \mathbf{x}_{k-1}}\}}{\mathrm{argmax}} \left|  \frac{dF}{d\varepsilon_{\mathbf{x}_i}}  \right|
\end{gathered}
\end{equation}
In Eq. \eqref{equation_selection}, the previously selected design features are excluded in the subsequent iterations of the algorithm. Our algorithm may contain multiple iterations of \eqref{equation_selection}, where each adjoint gradient profile is computed per iteration using only two (forward- and adjoint-) simulations from the adjoint sensitivity analysis. 

The material densities of the selected design features are then inverted according to the aforementioned rules. Specifically, if a design feature has a positive adjoint gradient and its current material has a lower permittivity, the material is transitioned to increase permittivity. However, if the gradient is positive and the design feature already consists of a high permittivity material, no action is taken on the design feature. When the adjoint gradient is negative, the material is switched to decrease the permittivity. The change in permittivity ($\Delta\varepsilon_{\textbf{x}_i}$) for each condition is described in Eq.~\eqref{equation_standard}.

% \begin{equation}
% \label{equation_standard}
% \Delta \varepsilon_{\mathbf{x}_i} = 
% \begin{cases} 
% \varepsilon_{\text{high}}-\varepsilon_{\text{low}} & \text{if } \frac{dF}{d\varepsilon_{\mathbf{x}_{i}}} >0 \text{ and } \varepsilon_{\mathbf{x}_i} = \varepsilon_{\text{low}} \\
% -(\varepsilon_{\text{high}}-\varepsilon_{\text{low}}) & \text{if } \frac{dF}{d\varepsilon_{\mathbf{x}_{i}}} <0 \text{ and } \varepsilon_{\mathbf{x}_i} = \varepsilon_{\text{high}}  \\
% 0 & \text{if } \frac{dF}{d\varepsilon_{\mathbf{x}_{i}}} >0 \text{ and } \varepsilon_{\mathbf{x}_i} = \varepsilon_{\text{high}} \\
% 0 & \text{if } \frac{dF}{d\varepsilon_{\mathbf{x}_{i}}} <0 \text{ and } \varepsilon_{\mathbf{x}_i} = \varepsilon_{\text{low}} 
% \end{cases} 
% \end{equation}

\begin{equation}
\label{equation_standard}
\Delta \varepsilon_{\mathbf{x}_i} = 
\begin{cases} 
\varepsilon_{\text{high}}-\varepsilon_{\text{low}} & \text{if } \frac{dF}{d\varepsilon_{\mathbf{x}_{i}}} >0 \text{ and } \varepsilon_{\mathbf{x}_i} = \varepsilon_{\text{low}} \\
-(\varepsilon_{\text{high}}-\varepsilon_{\text{low}}) & \text{if } \frac{dF}{d\varepsilon_{\mathbf{x}_{i}}} <0 \text{ and } \varepsilon_{\mathbf{x}_i} = \varepsilon_{\text{high}}  \\
0 & \text{otherwise}
\end{cases} 
\end{equation}
In Eq.~\eqref{equation_standard}, $\varepsilon_{\text{high}}$ denotes the permittivity of the material with a higher value, while $\varepsilon_{\text{low}}$ represents that of a lower value. In our multi-layer metalens design problem, $\varepsilon_{\text{high}}$ is equivalent to $\varepsilon_{\text{TiO$2$}}$ and $\varepsilon_{\text{low}}$ to $\varepsilon_{\text{SU-8}}$. After the inversion process, the FoM values of the newly generated devices are labeled using Eq.~\eqref{equation_change}.

\begin{align}
&F_1 = F_0 + \frac{dF}{d\varepsilon_{\mathbf{x}_{1}}}\Delta \varepsilon_{\mathbf{x}_1} \\
&F_2 = F_0 + \frac{dF}{d\varepsilon_{\mathbf{x}_{1}}}\Delta\varepsilon_{\mathbf{x}_1} + \frac{dF}{d\varepsilon_{\mathbf{x}_{2}}}\Delta \varepsilon_{\mathbf{x}_2}\label{equation_change}\\ 
& \hspace{25mm} \vdots \\
&F_k = F_0 + \sum_{i=0}^{k-1}\frac{dF}{d\varepsilon_{\mathbf{x}_{i}}}\Delta \varepsilon_{\mathbf{x}_i}
\end{align}
Here, we use $F_1$ to $F_k$ to construct a new dataset, excluding $F_0$. This approach results in a dataset expansion of $k-1$ times the original dataset size, which reduces simulation costs associated with data generation. In Eq.~\eqref{equation_change}, when $\Delta \varepsilon_{\mathbf{x}_i} = 0$, there is no alteration in the structure or in the value of the FoM in the step's progression.

\begin{figure}[H]
    \centering
    \centerline{\includegraphics[width =1\textwidth]{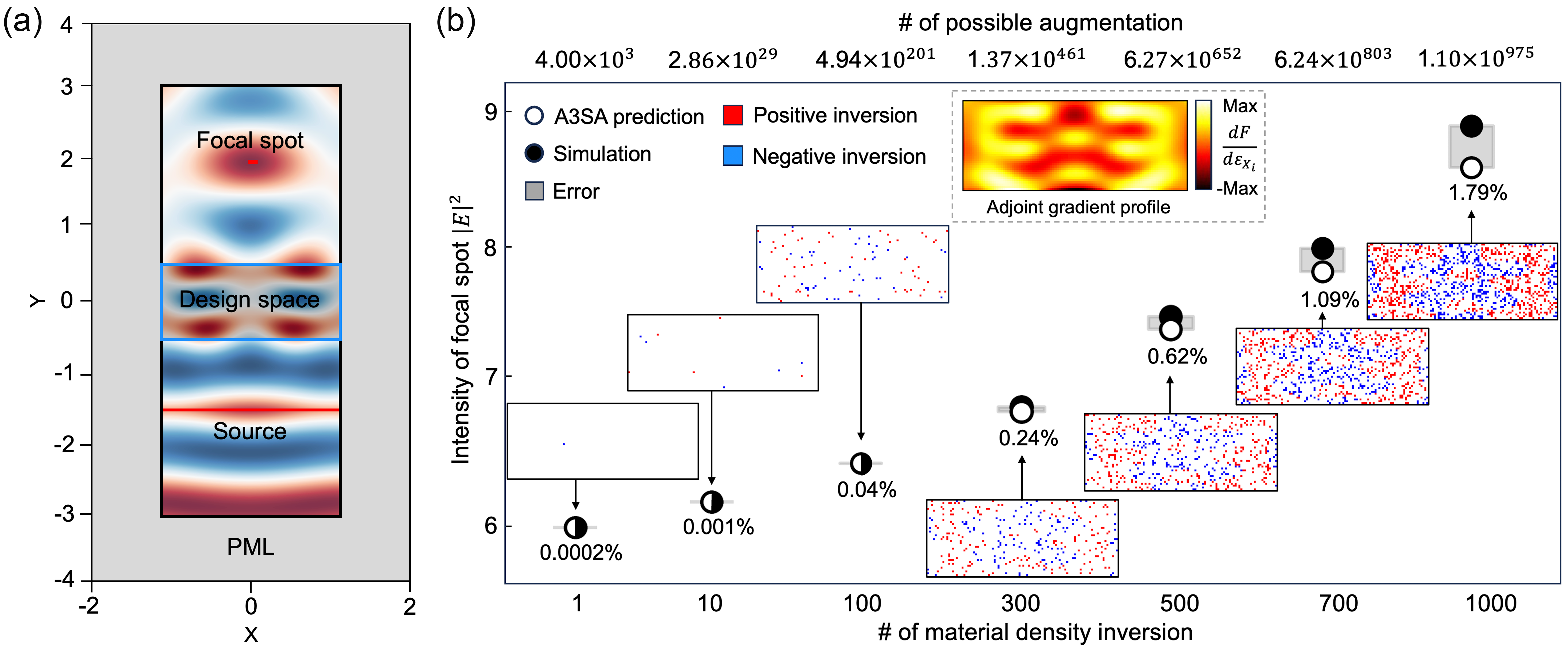}}
    \captionsetup{justification=raggedright}
    \caption{Validation of A3SA in a free-form 2D structure. Our validation begins with a planar 2D structure. We then compare the FoMs of the augmentation data to those of the simulated data (ground truth). (a) The simulation environment in a two-dimensional design space composed of 4,000 pixels. We set the electric field intensity at the focal point as the FoM in the optimization process. (b) The adjoint gradient profile is calculated via the adjoint sensitivity analysis, as shown in the inset surface plot. 
    }
\label{fig: Figure 2}
\end{figure}

To confirm the accuracy and improvement of data distribution of the A3SA, we perform a validation study in a free-form 2D structure. Our validation begins with a planar 2D structure with 4,000 pixels in the design space, as shown in Fig.~\ref{fig: Figure 2}(a). We assume a two-dimensional lens problem which is a field maximization at its focal point. First, the adjoint gradient profile is calculated by adjoint sensitivity analysis, as shown in the inset surface plot of Fig.~\ref{fig: Figure 2}(b). Based on this, we randomly select the locations of material density inversions for 1 to 1,000 pixels where negative inversion occurs for negative adjoint gradient while positive inversion occurs for positive adjoint gradient. Then, we obtain simulated FoMs (ground truth) for the augmented (inverted) photonic structures to calculate the error of the FoM prediction (denoted as gray boxes) of the A3SA. As shown in Fig.~\ref{fig: Figure 2}(b), the A3SA successfully predicts FoM changes over 500-pixel inversions over a total of 4,000 pixels with less than $1\%$ prediction error. Theoretically, this 500-pixel inversion corresponds to the possible data augmentation of 6.27$\times 10^{652}$ since we can randomly select the locations of the inversion within the 4000 pixels in the design space. This is an extraordinary data augmentation enabled by only two simulations. Also, 700 to 1,000-pixel inversions demonstrate prediction errors of 1.09\% and 1.79\%, respectively. Therefore, they can also be employed in the data augmentation of deep generative models such as variational autoencoders (VAE)~\cite{kingma2022autoencoding}, GAN~\cite{goodfellow2014generative}, or diffusion models~\cite{ho2020denoising, dhariwal2021diffusion}.

\section{Multilayered Metalenses}
\begin{figure}[H]
    \centering
    \captionsetup{justification=raggedright}

    %\centerline{
    \centerline{\includegraphics[width =\textwidth]{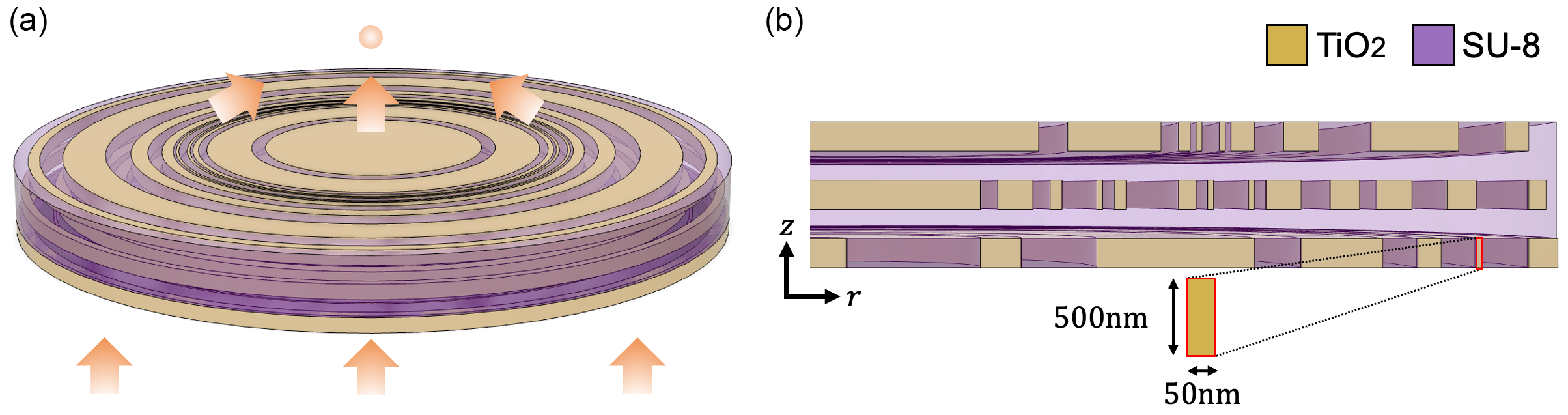}}

      \caption{Illustration of a multi-layer metalens design problem. (a) A circularly polarized plane wave propagates from the bottom of the cylindrically-symmetric metalens. The cross-sectional view of the metalens is depicted in (b). We confine the designable region to three layers of nanoring structures composed of TiO\textsubscript{2} and SU-8 materials. Intermediate layers are filled with SU-8 to make the multi-layer structure fabricable.}
    \label{fig: Figure 3}
\end{figure}
% 이 문제가 왜 중요한지
%metalens 소개
%Metasurface is a flat optical device with subwavelength structures that manipulate incidence waves in unprecedented ways~\cite{chen2016review, genevet2017recent}. This new approach enables a new way of compact imaging through a metalens, a two-dimensional device that focuses light with a geometric phase delay~\cite{khorasaninejad2017metalenses,chen2012dual,kang2012wave}. However, due to its limited interaction of light and components, improving the performance of metalenses has been of great interest in the domain of photonics. In this context, multi-layered metalens has emerged as next-generation metalens due to its superior controllability, higher efficiency, and increased degree of freedom in the manipulation of light~\cite{cheng_multi}.
% 다시 썼음 (해준)
Metasurfaces are flat optical devices with subwavelength structures that manipulate incident waves in unprecedented ways, providing a more remarkable precision of the manipulation than their bulky conventional counterparts~\cite{chen2016review, genevet2017recent}. This approach enables a new way of compact imaging through a metalens, a two-dimensional device that focuses light with a geometric phase delay~\cite{khorasaninejad2017metalenses,chen2012dual,kang2012wave}. However, the standard metalens design approach, which stitches subwavelength unit cells together into a larger device, is limited to low numerical apertures or low focusing efficiencies due to sampling errors in the stitching process~\cite{chung2020high, wang2018broadband, li2020dual, pan2022dielectric}. Recent studies~\cite{kamali2018review, presutti2020focusing} suggest that increasing the volume of the metasurface may relax the limitation of the metasurface's performance due to both increased geometric degrees of freedom and the provision of more room for light manipulation. However, a conventional metasurface design approach, known as unit-cell design, cannot provide a blueprint for multi-layer design because it cannot predict the interactions among the unit cells in different layers. Therefore, in this study, we apply our proposed data augmentation algorithm to solve the high-NA multi-layer metalens design problem.

% 우리가 이러한 메타렌즈 설계 문제로 타겟한 이유는 1. 공정 가능성, 2. 멀티 레이어 구조로 인한 많은 degree of freedom
We employ cylindrical symmetry in our designs to minimize the computational burden without sacrificing the focusing efficiency, as illustrated in Fig.~\ref{fig: Figure 3}(a). 
The Fraunhofer diffraction~\cite{born2013principles} from circular apertures results in a pattern called an Airy disk~\cite{airy1835diffraction}. This pattern features a dark region, referred to as a dark ring, where the destructive interference of light occurs. We define the focusing efficiency of our metalens by integrating focused energy within the third dark ring. The design space is confined to the r-z cross-section of the cylindrical metalenses, consisting of multi-layer TiO\textsubscript{2} nanostructures with SU-8 background, as illustrated in Fig.~\ref{fig: Figure 3}(b). TiO\textsubscript{2} and SU-8 offer a refractive-index difference of 0.9264 at a $1000$nm wavelength, making them suitable components for highly resonant nanophotonic structures~\cite{tio2}~\cite{su8}. The strong resonance is crucial for designing ``fast lenses" (high-NA), where a required phase profile rapidly varies over the radial direction of the lens. Moreover, a TiO\textsubscript{2} nanopattern with SU-8 background is feasible for fabrication by electron-beam lithography of 
SU-8, a commonly used epoxy-based negative photoresist used in microfabrication and spin-coating with SU-8~\cite{Mansouree:20}. The full-wave simulations are performed using Meep~\cite{OSKOOI2010687, Hammond:22}, an open-source software package for a finite-difference time-domain (FDTD) simulation. The minimum grid spacing of the FDTD simulation is $50$nm, which corresponds to the minimum width of the TiO\textsubscript{2} nanostructures in the multi-layer metalens. The design parameters are NA=0.75, wavelength=$1000$nm, focal length=5.65$\lambda$; we design the structure as shown in Fig.~\ref{fig: Figure 3} and use cylindrical symmetry to reduce the computational costs of the design process. In the problem setup of a cylindrical metalens, multiple pixels are clustered in a nanoring. This indicates that each inversion of the nanoring structure involves several pixels simultaneously. Each pixel may have different adjoint values; thus, the material density inversion predicted by a spatially averaged adjoint value in each nanoring may not increase FoM significantly unlike the inversion of a free-form structure. %It may lead to a higher error compared with the experiment in FIG.~\ref{fig: Figure 2}. This configuration hinders exact labeling. Nonetheless, the A3SA algorithm is sufficiently valid because of its capability to improve the photonic dataset without the need for labeling, which is demonstrated in subsequent experiments. 
We note that the multi-layered metalens setup is influenced by the findings presented in previous studies~\cite{mansouree2020multifunctional, chung2022inverse, doi:10.1021/acs.nanolett.2c02339, cheng_multi}.

\section{Results}
\begin{figure}[H]
    \centering
    \centerline{\includegraphics[width =0.5\textwidth]{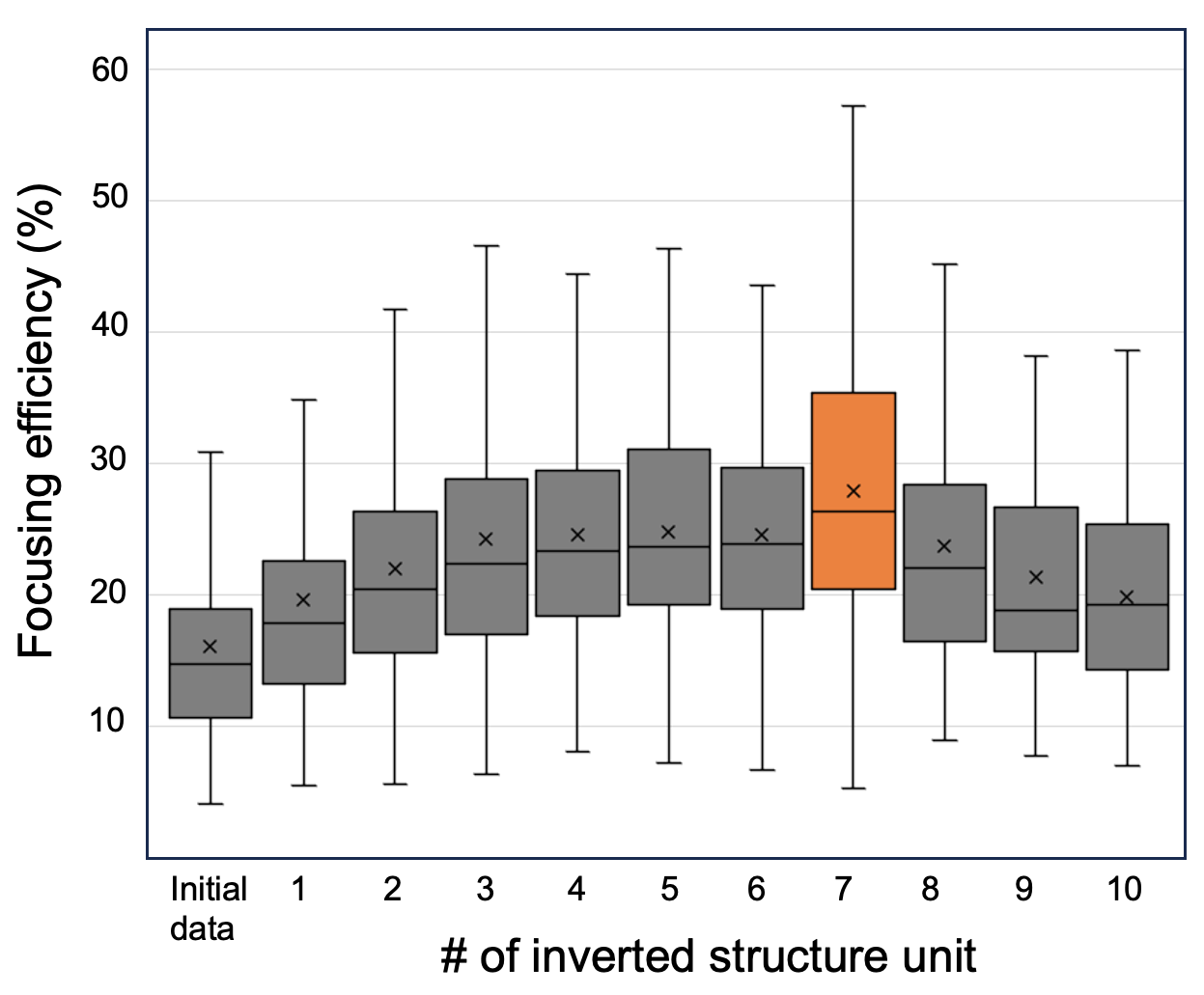}}
    \captionsetup{justification=raggedright}
    \caption{Box plot illustrating the focusing efficiency for randomly initialized multi-layer metalenses and A3SA-generated data derived from the initial dataset. Each plot provides a summary of each dataset, and the minimum and maximum values are denoted by the whiskers' lower and upper ends, respectively. The bottom and top edges of the box represent the first (Q1) and third (Q3) quantiles, respectively. The median value is indicated by the line within the box. The mean value is depicted as an ``x" mark inside the box. An inversion number of seven is chosen as it exhibits the peak average efficiency, with a maximum focusing efficiency value of 57.18\%, a mean of 28.36\%, and a standard deviation of 10.85\%.}
    \label{fig: Figure 4}
\end{figure}

First, we study the threshold of multiple inversions of the nanoring structure, which is equivalent to one iteration of A3SA, in the multi-layer metalens structure. The numerical experiment is motivated by the insight that a large number of inversions may result in breaking of the conditions of the Born approximation validity. As illustrated in Fig.~\ref{fig: Figure 4}, the experiment shows that the new FoMs gradually increase over a greater number of inversions up to seven and then decrease, which implies that the Born approximation may be violated around the seventh' inversions, leading to a failure of FoM prediction in the new structure.

% Figure 4

\begin{figure}[H]
    \centering
    \captionsetup{width=\textwidth, justification=raggedright}
    \includegraphics[width =\textwidth]
    {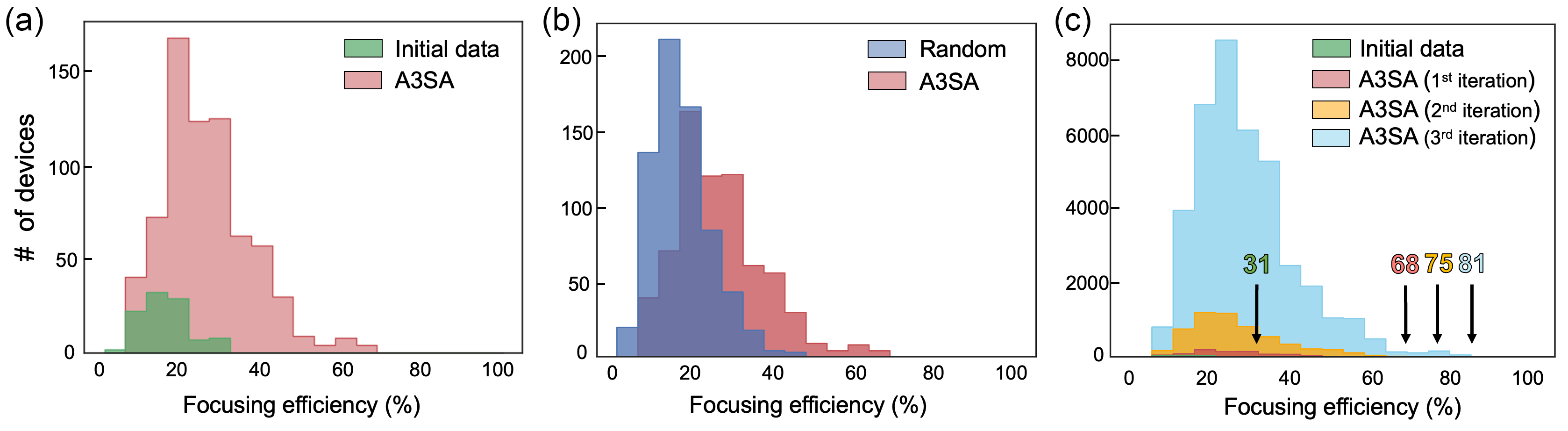}
    \caption{Analysis of the data distribution in the multiple iterations of the A3SA. (a) The initial data has a size of 100, where the average and maximum focusing efficiencies are 15.42\% and 31.05\%, respectively. After applying a single step of A3SA, the dataset size increases to 700. The distribution after the process has an average and maximum efficiency of 25.63\% and 68.24\%. (b) Seven hundred randomly generated data are compared against the A3SA-generated data. The average efficiency of the randomly generated data is 16.20\%, and the maximum efficiency is 43.38\%, which are lower than those of the A3SA-generated data  (c) Comparison of the datasets generated per each iteration of A3SA. After successive expansions of the A3SA process, the initial one hundred data grew to 700 in the first iteration, 4,900 in the second, and 32,300 in the third iteration. In parallel, the maximum focusing efficiency notably increases from 31.05\% (initial data) to 68.24\% (first iteration), 75.04\% (second iteration), and 81.39\% (third iteration). The sequential enhancement of the focusing efficiency and the number of augmented data proves the effectiveness of the algorithm.}
    \label{fig: Figure 5} 
\end{figure}

We apply multiple iterations of the A3SA to the multi-layer metalens design problem to demonstrate the efficacy of this method. Specifically, we start with one hundred randomly generated initial data entries (green bars) shown in Fig.~\ref{fig: Figure 5}(a), where they have an average focusing efficiency of 15.42\% and a maximum focusing efficiency of 31.05\%. Then, the A3SA augments one hundred data to seven hundred by multiple inversions of the material density of the nanoring structure. The data distribution after the single iteration of the A3SA shows an average and maximum efficiency of 25.63\% and 68.24\%, respectively, as illustrated in Fig.~\ref{fig: Figure 5}(a). The second iteration of the A3SA is applied to the 700 hundred data obtained from the first iteration. The total amount of data is now 4,900 in the second and 32,300 in the third iteration. The maximum focusing efficiency increases significantly over the multiple iterations of the A3SA. It ranges from 31.05\% (initial data) to 68.24\% (first iteration), 75.04\% (second iteration), and 81.39\% (third iteration). The sequential enhancement of the focusing efficiency and the number of augmented data sets proves the effectiveness of the algorithm. We also compare the augmented data with randomly generated data with the same amount as shown in Fig.~\ref{fig: Figure 5}(b). The augmented data shows higher average (25.63\%) and maximum (68.24\%) efficiencies compared to the randomly generated data.

%multiple A3SA iterations to multi-layer metalenses and investigate the distribution of the augmented data. FIGURE~\ref{fig: Figure 5}(a) shows the randomly generated initial data (green) versus the newly generated data (red) obtained through the A3SA process. The initial data has a size of 100, and the average and maximum focusing efficiencies are 15.42\% and 31.05\%, respectively. After applying A3SA, the dataset size is increased to 700 by multiple inversions of the material density of the nanoring structure. The distribution after the A3SA process has an average and maximum efficiency of 25.63\% and 68.24\%, respectively. FIGURE~\ref{fig: Figure 5}(b) compares the data generated by the A3SA and a set of randomly created devices that are equal in number. The augmented data in FIG.~\ref{fig: Figure 5}(b) shows higher average (x\%) and maximum (y\%) efficiencies. When the A3SA process is repeated multiple times, the augmented data distribution exhibits higher focusing efficiencies. Specifically, the dataset undergoes successive expansions: starting from 100, it increases to 700 in the first iteration, 4,900 in the second, and finally to 32,300 in the third iteration. The maximum focusing efficiency increases significantly. It ranges from 31.05\% (initial data) to 68.24\% (first iteration), 75.04\% (second iteration), and 81.39\% (third iteration). The sequential enhancement of the focusing efficiency proves the effectiveness of the algorithm in optimizing the FoM distribution. 

\begin{figure}[H]
    \centering
    \centerline{\includegraphics[width =0.95\columnwidth]{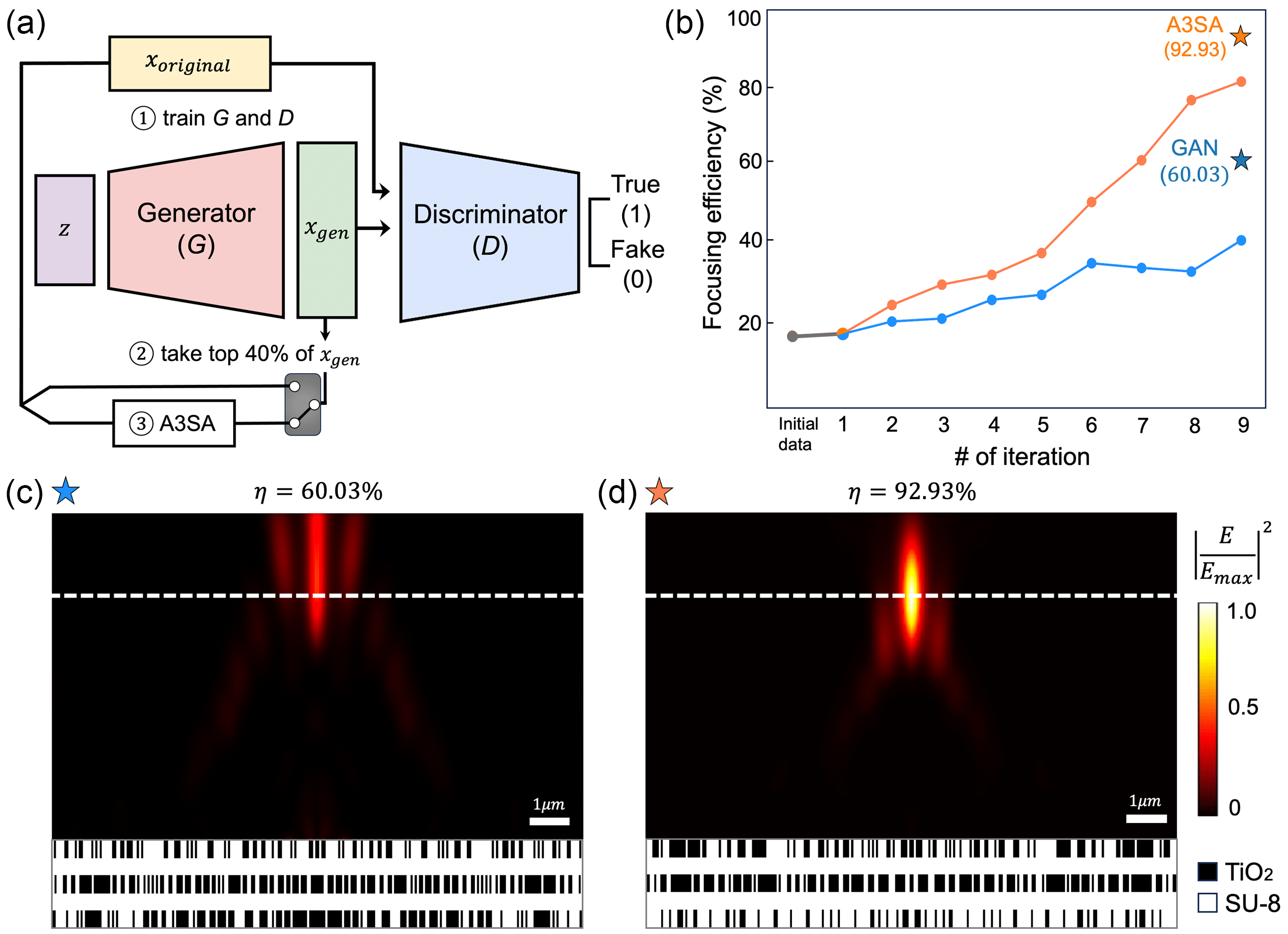}}
    \caption{(a) Schematics of the integration of A3SA with GAN. $\textcircled{\small{1}}$ We train both the generator ($G$) and discriminator ($D$), which constitute the GAN. The generator ($G$) synthesizes fake data ($x_{gen}$), and the discriminator ($D$) distinguishes if the data are from the true dataset ($x_{original}$) or the generator. $\textcircled{\small{2}}$ In each iteration, we select the top 40\% of the devices generated from the generator. $\textcircled{\small{3}}$ We apply A3SA to enhance the input dataset, which is then utilized as training data ($x_{original}$) for the subsequent iteration of the process. An ablation setup is also illustrated as an upper circuit of the switch, which only involves $\textcircled{\small{1}}$ and $\textcircled{\small{2}}$. (b) Result of a comparative analysis performed between the A3SA algorithm (marked as A3SA) and its corresponding ablation study (marked as GAN). The dotted lines illustrate the average focusing efficiency of a dataset for each iteration of (a). At the final (ninth) iteration, A3SA displays 81.92\% and GAN shows 41.57\% of average focusing efficiency. (c) and (d) Illustration of the normalized field intensity profile of the optimized multi-layer metalenses by `GAN' (c) and `A3SA' (d), respectively. A white dotted line denotes the desired focal length. The field intensity profile in (d) validates the high focusing efficiency (92.93\%) of the metalens structure generated by A3SA. The black (TiO$_2$) and white (SU-8) structures indicate a cross-section of the optimized metalenses.}
    \label{fig: Figure 6}
\end{figure}

We also implemented a generative adversarial network (GAN) approach to design multi-layer metalenses using A3SA data. GANs are renowned for their fast inference speeds, and they have been already successfully implemented in the inverse design of photonic structures~\cite{jiang2019free, an2021multifunctional, 10.1145/3447548.3467414}. Figure~\ref{fig: Figure 6}(a) illustrates the schematics of the GAN based on the A3SA data. The generative model used in this study comprises two networks: a generator ($G$) and a discriminator ($D$). The generator generates structural data $x_{gen} = G(z)$ from the random input noise $z$. The discriminator determines whether the input data are fake (labeled 0) or true (labeled 1). In the $\textcircled{\small{\textbf{1}}}$ training process illustrated in Fig.~\ref{fig: Figure 6}(a), we train two networks by adversarial learning grounded in the minimax optimization of a loss function $L (D, G)$, which is mathematically expressed in Eq.~\eqref{GANloss}.

\begin{equation}
\min_G \max_D L(D, G)=\mathbb{E}_{x \sim p_{\text{data}}(x)}[\log D(x)]+\mathbb{E}_{z \sim p_{z}(z)}[\log (1-D(G({z})))] \label{GANloss}
\end{equation}

As described in Eq.~\eqref{GANloss}, generator $G$ aims to minimize the loss, and discriminator $D$ attempts to maximize the loss simultaneously. Note that in the first term, data sampled from the true dataset $p_{\text {data}}(x)$ are employed to train the discriminator $D$, whereas in the second term, the generated data $G(z)$ are used to train both the generator $G$ and the discriminator $D$. During the training process, a continuous interplay occurs between the generator and the discriminator. The generator attempts to deceive the discriminator by creating indistinguishable synthetic samples. In contrast, the discriminator attempts to distinguish true samples from the synthetic samples generated by the generator. The iterative process continues until the two networks reach a Nash equilibrium~\cite{nash1950equilibrium, nash1951non}, where the generator is expected to produce high-quality samples that the discriminator can hardly differentiate from the true samples.

A progressive enhancement of data generation can be achieved by successive iterations of stages $\textcircled{\small{\text{1}}}$, $\textcircled{\small{\text{2}}}$, and $\textcircled{\small{\text{3}}}$, as depicted in Fig.~\ref{fig: Figure 6} (a). In $\textcircled{\small{\text{1}}}$, we train both the generator ($G$) and the discriminator ($D$) networks together. This is succeeded by $\textcircled{\small{\text{2}}}$, where we take the top 40th percentile of generated devices from the trained generator, which is a strategy benchmarked from the previous study~\cite{jiang2019free}. The filtered devices are then used as training data for the next iteration of the process. The GAN provides stochasticity to the data distribution, avoiding the convergence to a bad local optima. To further enhance the filtered data, we additionally apply $\textcircled{\small{\text{3}}}$ A3SA to them. In $\textcircled{\small{\text{3}}}$, we utilize a subset of the data generated from the A3SA, ensuring that the size of both input and output data of the A3SA remains constant. This ensures fairness of the comparative study between A3SA-based GAN and basic GAN.

A comparison of the A3SA algorithm (labeled A3SA) with its corresponding ablation study (labeled GAN) is shown in Fig.~\ref{fig: Figure 6} (b). In the ablation setup, the proposed A3SA process is excluded so that only $\textcircled{\small{\text{1}}}$ and $\textcircled{\small{\text{2}}}$ in Fig.~\ref{fig: Figure 6} (a) are performed per iteration. Both A3SA-based GAN and basic GAN demonstrate increased focusing efficiencies over the iteration. However, the maximum and average focusing efficiencies of the A3SA-based GAN are much higher than those of the basic GAN. Specifically, at the ninth iteration of the A3SA employed GAN, we find a multi-layer metalens design that demonstrates 92.93 \% focusing efficiency, which is close to the theoretical maximum efficiency ($\sim$93.80\%) of the third dark ring of the Airy disk~\cite{born2013principles, airy1835diffraction}. Figures~\ref{fig: Figure 6} (c) and (d) show the normalized field intensities of the optimal structures discovered from the GAN and the combination of the GAN with the A3SA, respectively. The intensity plot in Fig.~\ref{fig: Figure 6} (c) corresponds to a structure with a 60.03\% efficiency, while Fig.~\ref{fig: Figure 6} (d) corresponds to a 92.93\% efficiency. At the target focal length indicated by a white dashed line, it is observable that the incident wave is more effectively focused in the A3SA-optimized multi-layer metalens. It implies that the A3SA combined with a machine-learning algorithm may pave a new way of designing ultra-high-efficiency photonic devices within a feasible amount of the simulations.

%To validate whether the FoM setting is performed appropriately, we also present the field simulation results. Fig.~\ref{fig: Figure 6} (c) and (d) show the normalized field intensities of the optimal structures discovered from the GAN and the combination of the GAN with the A3SA, respectively. The field plot in (c) corresponds to a structure with a 60.03\% efficiency, while (d) corresponds to a 92.93\% efficiency. At the target focal length indicated by a white dotted line, it is observable that the light is more effectively focused in (d). The results confirm our FoM setting for the focusing efficiency, as well as the validity of the A3SA.

\section{Conclusion}
 In this work, we have demonstrated a novel way of augmenting photonic device designs without running numerious simulations. The proposed A3SA algorithm is built on the principle of adjoint sensitivity analysis, forecasting changes in the figure of merit resulting from structural perturbations. By leveraging the gradient values, we can augment and label numerous new designs without additional computations. We validate the A3SA both in free-form design and multi-layer metalens design problems. In the former example, A3SA successfully generates new data within $1\%$ prediction error and shows the possible data augmentation of 6.27$\times 10^{652}$. In the multi-layer design problem, it has demonstrated a capability of generating more than 300 times the amount of original data while enhancing the device efficiency. After incorporating the proposed algorithm into a GAN, the optimized metalens exhibits a maximum focusing efficiency of 92.93\%, comparable to the theoretical upper bound (93.80\%). Our method opens a promising way of sidestepping major hurdles, data generation and initial data distribution, of using deep learning in photonics.

\nocite{*}

\bibliography{aps}% Produces the bibliography via BibTeX.

\end{document}